\documentclass[prl,preprintnumbers,twocolumn,showpacs,showkeys,superscriptaddress,citeautoscript]{revtex4-1}

\usepackage{amsmath,amssymb}

\usepackage[colorlinks=true,citecolor=blue,linkcolor=blue]{hyperref}
\usepackage{physics}

\usepackage{acronym}

\usepackage{todonotes}
\usepackage{lipsum}
\usepackage{microtype}

\begin{document}
\preprint{CP3-Origins-2019-35 DNRF90}

% * title page
\title{%Mini review of the large $N_f$ gauge-fermion theories \\ and \\ 
Comment on `` A critical look at $\beta$-function singularities at large $N$" \\ by \\  Alanne, Blasi and Dondi}

 \author{Francesco Sannino}
\email{sannino@cp3.sdu.dk}
\author{Zhi-Wei Wang}
\email{wang@cp3.sdu.dk}
\affiliation{CP3-Origins \& the Danish Institute for Advanced Study. University of Southern Denmark. Campusvej 55, DK-5230 Odense, Denmark}

\begin{abstract}
We first briefly review the state-of-the-art of the large $N_f$ gauge-fermion theories and then show that the claim made in \cite{Alanne:2019vuk}  that ``The singularities in the $\beta$ function and in the fermion mass anomalous dimension are simultaneously removed providing no hint for a UV fixed point in the large-N limit" is unwarranted. This is so since the author's  truncation of the beta function violates the large number of matter fields counting.   \end{abstract}
\maketitle

% these have to come after the maketitle for the acronyms to work
\acrodef{vev}{vacuum expectation value}
\acrodef{cft}[\textsc{cft}]{conformal field theory}
\acrodefplural{cft}[\textsc{cft}s]{conformal field theories}
\acrodef{eft}[\textsc{eft}]{effective field theory}
\acrodef{nlsm}[\textsc{nlsm}]{non-linear sigma model}

%%%%%%%%%%%%%%%%%%%%%%%%%%%%%%%%%%%%
% * introduction
%%%%%%%%%%%%%%%%%%%%%%%%%%%%%%%%%%%%
The discovery of  asymptotic safety in \cite{Litim:2014uca}  in perturbatively controllable gauge-Yukawa theories  has spurred  interest in understanding the ultraviolet fate of other non asymptotically free theories. An interesting class is constituted by gauge-fermion theories at large number of fermion matter fields and fixed number of colors. QED and QCD at large number of fermion fields are two noteworthy examples. Differently from  \cite{Alanne:2019vuk} here we indicate with $N_f$ the number of matter fields to avoid potential confusion with $N$ which is often taken to be the number of colors of the theory. Here too we will take the number of colors to be $N$. 

The first non-trivial order in the  $1/N_f$ expansion for the beta function and anomalous dimension of these theories were computed long time ago in \cite{PalanquesMestre:1983zy,Gracey:1996he}. The physical consequences of these beta functions were first discussed and critically analysed in \cite{Holdom:2010qs} and \cite{Pica:2010xq} and later in \cite{Shrock:2013cca}.  To this order the beta function develops a singularity that differs for the abelian and non-abelian groups as summarised in  \cite{Antipin:2017ebo}. It turns out that the singular structure, at order $1/N_f$, induces a zero in the beta function. Of course, the zero  in the beta function must be taken with the grain of salt since it is a result of a delicate cancellation between the leading order and a subleading order in $1/N_f$. This caused  excitement that QED could indeed have a UV fixed point at leading order in $1/N_f$. However,   in  \cite{Antipin:2017ebo} it was shown that the fermion anomalous dimension of the fermion quark mass operator at the alleged fixed point explodes.  This causes the operator to violate the unitarity bound. Therefore either the fixed point is unphysical or  the fermion mass operator decouples. Since fermions are crucial for the existence of the fixed point it is hard to believe that this operator decouples. Therefore for QED the leading $1/N_f$ fixed point is not expected to be physical. Nota bene,  this result does not imply that QED cannot have a fixed point in the full theory. 

On the other hand, for  non-abelian gauge-fermion theories, the anomalous dimension of the fermion mass operator drops as  $1/N_f$ without violating the unitarity bound.  Additionally, in a recent paper \cite{Ryttov:2019aux}  the baryon and glueball anomalous dimensions were computed and shown that only the glueball anomalous dimension violates the unitarity bound.   Therefore the glueballs states decouple at the non-abelian UV potential fixed point. This, per se, is physically understandable since the dynamics is due to the fermions and not to the gluons in the infinite number of fermion limit. 

In an attempt to go beyond the leading $1/N_f$ computation the authors of \cite{Alanne:2019vuk}   used the relation between the beta function in four dimensions and the critical exponent    $\omega(d) = \sum_{n=0}^{\infty} \frac{\omega^{(n)}}{N_f^n}$ of the Wilson-Fisher fixed point in less than four dimensions.   In the table below we summarise when, in the $1/N_f$ expansion, each critical exponent starts contributing.  
 \begin{center}
\begin{table} [h!]
\scalebox{1.3}{%
\begin{tabular}{c|c|c|c|c|}
%  \cline{1-5}
   & $\omega^{(1)}$ & $\omega^{(2)}$  & $\omega^{(3)}$  & $\cdots$ \\ \cline{1-5}
  $1/N_f$ & $F_1^{(1)}$     \\ \cline{1-3}
 $1/N_f^2$ & $F_2^{(1)}$ & $F_2^{(2)}$  \\ \cline{1-4}
 $1/N_f^3$ & $F_3^{(1)}$ & $F_3^{(2)}$ &  $F_3^{(3)}$  \\ \cline{1-5}
 $\cdots$ & $\cdots$ & $\cdots$ & $\cdots$ & $\cdots$\\ \cline{1-4}\hline
\end{tabular}}
\caption{Order  in the $1/N_f$ expansion when each $\omega^{(n)}$ term starts contributing for the large $N_f$ beta function.  }
\end{table}
\end{center}
Schematically the  beta function in $K  \propto \alpha N_f$ reads: 
\begin{equation}
\frac{\beta(K)}{K^2} = 1 + \frac{F^{(1)}_1}{N_f} + \frac{F^{(1)}_2 + F^{(2)}_2}{N_f^2} + \frac{F^{(1)}_3+F^{(2)}_3 + F^{(3)}_3}{N_f^3} + \cdots \ ,
\label{beta}
\end{equation}
with  $F^{(j)}_n (K)$ a non-asymptotic function of $K$. 
Obviously, if all contributions are taken into account the full beta function can be computed which is independent of the large $N_f$ limit. The resulting series in $K$ will be asymptotic, and up to instanton contributions, will yield the exact result. However so far we only know $\omega^{(1)}$ \cite{Gracey:1996he} and therefore a consistent computation in $1/N_f$ cannot be performed. Nevertheless in \cite{Alanne:2019vuk} the authors obtained a truncated version of the full beta function by resumming  the contributions stemming from the first column of the table that it is due indeed only to $\omega^{(1)}$. The truncated beta of function of \cite{Alanne:2019vuk}  has therefore the following issues:
\begin{itemize}
\item[i)]{the truncation breaks the large $N_f$ counting,}
\item[ii)]{because of i) the radius of convergence of each term in the $1/N_f$ expansion is incorrect,}
\item[iii)]{The (non)singular structure of the truncated beta function does not reflect the (non)singular structure of the full theory nor of each term in the $1/N_f$ expansion. }
\end{itemize}
  In fact, as shown in \cite{Dondi:2019ivp}, the correct analytic structure of each term in the $1/N_f$ expansion can be determined by expanding each numerator of \eqref{beta} in powers of $K$. For the leading term in $1/N_f$ one needs at least 30 perturbative orders. Clearly a different radius of convergence and a different analytic structure would have emerged already at leading order in $1/N_f$ if we would have chosen a partial subgroup of terms in the sum that is not organised according to a well defined limit.  

%The missing $\omega^{(n))}$ with $n>1$  contributions to the beta function can, at each order and starting already  at the order $1/N_f^2$,   reinstatate the old singularity or add new ones.   

Of course, this is a hard subject and therefore we very much  appreciate the efforts made by the authors in  \cite{Alanne:2019vuk} to go beyond the state-of-the art.

We summarise our comment by re-iterating that from the truncated beta function of \cite{Alanne:2019vuk} no firm statement can be made about the (non)singular structure of  large $N_f$  gauge-fermion theories. Lattice simulations, on the other hand,  will be able to test their UV nature   \cite{Leino:2019qwk}. 

\bigskip
\acknowledgments
We thank Tommi Alanne, Simone Blasi and Nicola Dondi for explaining in detail their work to us. This work is partially supported by the Danish National Research Foundation grant DNRF:90.


\begin{thebibliography}{99}
% \newpage
%\printbibliography{}
%\cite{Alanne:2019vuk}
\bibitem{Alanne:2019vuk} 
  T.~Alanne, S.~Blasi and N.~A.~Dondi,
  %``A critical look at $\beta$-function singularities at large $N$,''
  arXiv:1905.08709 [hep-th].
  %%CITATION = ARXIV:1905.08709;%%
  %2 citations counted in INSPIRE as of 15 Sep 2019
  
%\cite{Litim:2014uca}
\bibitem{Litim:2014uca} 
  D.~F.~Litim and F.~Sannino,
  %``Asymptotic safety guaranteed,''
  JHEP {\bf 1412}, 178 (2014)
 % doi:10.1007/JHEP12(2014)178
  [arXiv:1406.2337 [hep-th]].
  %%CITATION = doi:10.1007/JHEP12(2014)178;%%
  %137 citations counted in INSPIRE as of 15 Sep 2019



%\cite{PalanquesMestre:1983zy}
\bibitem{PalanquesMestre:1983zy} 
  A.~Palanques-Mestre and P.~Pascual,
  %``The 1/$N^-$f Expansion of the $\gamma$ and Beta Functions in {QED},''
  Commun.\ Math.\ Phys.\  {\bf 95}, 277 (1984).
%  doi:10.1007/BF01212398
  %%CITATION = doi:10.1007/BF01212398;%%
  %99 citations counted in INSPIRE as of 15 Sep 2019

%\cite{Gracey:1996he}
\bibitem{Gracey:1996he} 
  J.~A.~Gracey,
  %``The QCD Beta function at O(1/N(f)),''
  Phys.\ Lett.\ B {\bf 373}, 178 (1996)
 % doi:10.1016/0370-2693(96)00105-0
  [hep-ph/9602214].
  %%CITATION = doi:10.1016/0370-2693(96)00105-0;%%
  %88 citations counted in INSPIRE as of 15 Sep 2019
%%% Local Variables:
%%% mode: latex
%%% TeX-master: t
%%% End:

%\cite{Holdom:2010qs}
\bibitem{Holdom:2010qs} 
  B.~Holdom,
  %``Large N flavor beta-functions: a recap,''
  Phys.\ Lett.\ B {\bf 694}, 74 (2010)
 % doi:10.1016/j.physletb.2010.09.037
  %[arXiv:1006.2119 [hep-ph]].
  %%CITATION = doi:10.1016/j.physletb.2010.09.037;%%
  %44 citations counted in INSPIRE as of 14 Feb 2019

%\cite{Pica:2010xq}
\bibitem{Pica:2010xq}
  C.~Pica and F.~Sannino,
  %``UV and IR Zeros of Gauge Theories at The Four Loop Order and Beyond,''
  Phys.\ Rev.\ D {\bf 83} (2011) 035013
%  doi:10.1103/PhysRevD.83.035013
  [arXiv:1011.5917 [hep-ph]].
  %%CITATION = doi:10.1103/PhysRevD.83.035013;%%
  %123 citations counted in INSPIRE as of 16 Sep 2019

%\cite{Shrock:2013cca}
\bibitem{Shrock:2013cca}
  R.~Shrock,
  %``Study of Possible Ultraviolet Zero of the Beta Function in Gauge Theories with Many Fermions,''
  Phys.\ Rev.\ D {\bf 89} (2014) no.4,  045019
  %doi:10.1103/PhysRevD.89.045019
  [arXiv:1311.5268 [hep-th]].


%\cite{Antipin:2017ebo}
\bibitem{Antipin:2017ebo} 
  O.~Antipin and F.~Sannino,
  %``Conformal Window 2.0: The large $N_f$ safe story,''
  Phys.\ Rev.\ D {\bf 97}, no. 11, 116007 (2018)
 % doi:10.1103/PhysRevD.97.116007
  [arXiv:1709.02354 [hep-ph]].
  %%CITATION = doi:10.1103/PhysRevD.97.116007;%%
  %26 citations counted in INSPIRE as of 15 Sep 2019
  
  %\cite{Ryttov:2019aux}
\bibitem{Ryttov:2019aux} 
  T.~A.~Ryttov and K.~Tuominen,
  %``Safe Glueballs and Baryons,''
  JHEP {\bf 1904}, 173 (2019)
  %doi:10.1007/JHEP04(2019)173
  [arXiv:1903.09089 [hep-th]].
  %%CITATION = doi:10.1007/JHEP04(2019)173;%%
  %3 citations counted in INSPIRE as of 15 Sep 2019
  
%\cite{Gracey:1996he}
%\bibitem{Gracey:1996he}
 % J.~A.~Gracey,
  %``The QCD Beta function at O(1/N(f)),''
 % Phys.\ Lett.\ B {\bf 373} (1996) 178
%  doi:10.1016/0370-2693(96)00105-0
 % [hep-ph/9602214].

  %\cite{Dondi:2019ivp}
\bibitem{Dondi:2019ivp} 
  N.~A.~Dondi, G.~V.~Dunne, M.~Reichert and F.~Sannino,
  %``Analytic Coupling Structure of Large $N_f$ (Super) QED and QCD,''
  Phys.\ Rev.\ D {\bf 100}, no. 1, 015013 (2019)
 % doi:10.1103/PhysRevD.100.015013
  [arXiv:1903.02568 [hep-th]].
  %%CITATION = doi:10.1103/PhysRevD.100.015013;%%
  %3 citations counted in INSPIRE as of 15 Sep 2019
  
  %\cite{Leino:2019qwk}
\bibitem{Leino:2019qwk} 
  V.~Leino, T.~Rindlisbacher, K.~Rummukainen, F.~Sannino and K.~Tuominen,
  %``Safety versus triviality on the lattice,''
  arXiv:1908.04605 [hep-lat].
  %%CITATION = ARXIV:1908.04605;%%
 \end{thebibliography}
  \end{document}